\documentclass[a4paper]{article}

\usepackage{INTERSPEECH2022}

\title{The ReturnZero System for VoxCeleb Speaker Recognition Challenge 2022}
\name{Sangwon Suh$^1$, Sunjong Park$^1$}

\address{
  $^1$Return Zero Inc., Seoul, Korea}
\email{simon@rtzr.ai, ryan@rtzr.ai}

\begin{document}

\maketitle
\begin{abstract}

  In this paper, we describe the top-scoring submissions for team RTZR VoxCeleb Speaker Recognition Challenge 2022 (VoxSRC-22) in the closed dataset, speaker verification Track 1. The top performed system is a fusion of 7 models, which contains 3 different types of model architectures. We focus on training models to learn extra-temporal information. Therefore, all models were trained with 4-6 second frames for each utterance. Also, we apply the Large Margin Fine-tuning \cite{liu2019large} strategy which has shown good performance on the previous challenges for some of our fusion models. While the evaluation process, we apply the scoring methods with adaptive symmetric normalization (AS-Norm) and matrix score average (MSA). Finally, we mix up models with logistic regression to fuse all the trained models. The final submission achieves 0.165 DCF and 2.912\% EER on the VoxSRC22 test set. 
  
\end{abstract}

\section{Introduction}

The Voxceleb Speaker Recognition Challenge (VoxSRC) has been held to develop state-of-the-art methods for recognizing speakers. This year's evaluation includes challenging trials for real-world situations such as background noise, laughter, reverbs, or any effects for recording environments. As same in previous years, VoxSRC-22 has four tracks of challenges, two tracks for speaker verification with a different range of datasets, a semi-supervised verification task, and speaker diarization.

For the speaker verification tracks, a recent trend is the enhancement of neural architectures for speaker embeddings. Many previous studies verified Convolution Neural Networks (CNNs) with attention pooling as a good structure for extracting speaker-related vectors. For example, ECAPA-TDNN \cite{desplanques2020ecapa} and RawNet \cite{jung2019rawnet, jung2020improved, jung2022pushing} consist of 1D convolutions, while ResNet \cite{chung2020defence, kwon2021ins, thienpondt2020idlab} and RepVGG \cite{ma2021rep, zhao2021speakin} are 2D. The superiority and inferiority of the two structural differences are not yet clear, but a significant improvement has been reported in ensembling each architecture's predictions.

This report describes the submitted systems for the VoxSRC-22 Track 1. We trained 3 different architectures and developed 7 systems combined with AS-Norm and MSA scoring methods. Among the model architectures, we included ResNet with a modification of stride configuration to enhance feature extraction with more temporal information.

\section{System descriptions}

\subsection{Datasets}

    The models are trained on the development set of VoxCeleb2 for Track 1, which contains 5,994 speakers and 1,092,009 utterances. In addition, we used the previous year's development dataset to evaluate the robustness of our models for different data distributions. The details of each dataset are as follows:

    \begin{itemize}
    \item \textbf{Vox1-O (cleaned)} : 37,611 trials sampled from VoxCeleb1 test dataset with 40 speakers
    \item \textbf{VoxSRC20-dev} : 263,486 trials that used for validation data in VoxSRC2020. This dataset consists of VoxCeleb1 dev and test dataset. Additionally, the same identities but out-of-domain VoxCeleb1 were included for this dataset.
    \item \textbf{VoxSRC21-val} : 60,000 trials in VoxCeleb1 domain datasets. This dataset focus on multi-lingual data.
    \item \textbf{VoxSRC22-dev} : 305,196 trials in VoxCeleb1 and VoxConverse domain datasets. This dataset focus on both the impact of age of the speaker on their speech segments and also the impact of shared background noise between speech segments from different speakers.
    \end{itemize}

\subsection{Data augmentations}

\subsubsection{Offline data augmentations}

    In general, the discrimination power of speaker embeddings improves when a larger number of speaker classes are learned. Therefore, we used speed augmentation \cite{ko2015audio} to increase the base size of classes. We used the SoX toolkit of torchaudio to apply speed perturbation with factors of 0.9, 1.0, and 1.1 to all utterances. As a result, we trained the model with 17,982 classes with 3,276,027 utterances.

\subsubsection{Online data augmentations}

    For robustness against various environments, it is beneficial to include noisy and reverberant audio in the training data. To achieve this, we followed the idea of \cite{kwon2021ins} which utilized the Kaldi recipes in an online manner. We used the audio source from MUSAN corpus \cite{snyder2015musan} and the impulse responses of small and medium rooms of RIRs noises dataset \cite{ko2017rir} to implement augmentation methods as follows:

    \begin{itemize}
    \item \textbf{AddNoise}: Add a noise source of 0-15dB SNR with a probability of 0.2
    \item \textbf{AddMusic}: Add a music source of 5-15dB SNR with a probability of 0.2
    \item \textbf{AddBabble}: Add a babbling noise, a summed signal of three to seven random speakers, of 13-20dB SNR with a probability of 0.2
    \item \textbf{AddReverb}: Add reverberated audio with a probability of 0.2
    \end{itemize}    

\subsection{ResNet-SE}
    
    The neural architecture of all systems utilizes the ResNet \cite{he2016deep} with the addition of squeeze-and-excitation blocks \cite{hu2018squeeze}. The ResNet-SE architecture is a competitive architecture in the speaker verification domain and has been used by the winning team of the VoxSRC21 \cite{zhao2021speakin}. Our implementation follows the 34-layer and 101-layer models, in which the number of convolution filters starts with 64.
    
    Furthermore, we assumed that only some frames of the utterance would have speaker-related information and expected that these frames might be lost by the aggressive compression on the time axis. For example, ECAPA-TDNN, widely used in the speaker recognition domain, does not pool feature maps during the convolution process. Therefore, we modified the stride configurations for each stage of the ResNet-SE-34 to apply this concept, as shown in Table \ref{tab:resnet_list}. In this new architecture, the frequency compression was configured conventionally, while the time stride was placed only once in the four residual stages. Although this method gave better results than the conventional ResNet-SE-34, it was impossible to redesign the 101 architecture due to excessive computation.
    
    After the convolution blocks, the feature map is summarized through attentive statistics pooling \cite{okabe2018attentive} and converted into 512-dimensional speaker embeddings. The input features of all models are the 80 log-scaled Mel-filterbanks with a cepstral mean normalization.
        
    \begin{table}[th]
      \caption{List of ResNet-SE variants}
      \label{tab:resnet_list}
      \centering
      \begin{tabular}{lcc}
        \toprule
        \textbf{Variants name} & \textbf{Backbone} & \textbf{Strides} \\
        \midrule
        \textbf{ResNet34-st1112} & ResNet-SE-34 & 1, (2,1), (2,1), 2 \\
        \textbf{ResNet34-st1121} & ResNet-SE-34 & 1, (2,1), 2, (2,1) \\
        \textbf{ResNet101} & ResNet-SE-101 & 1, 2, 2, 2 \\
        \bottomrule
      \end{tabular}
    \end{table}

\subsection{Subcenter additive angular margin loss}

    The original softmax loss function which is commonly used for classification tasks represented as follows:

    \begin{equation}\label{eq1}
        L(x_i) =  - \log ( \frac{e^{W^T_{y_{i}}x_i + b_{y_i}}}{e^{W^T_{y_{i}}x_i + b_{y_i}}})
    \end{equation}

    The main problem for the  original softmax loss function is that it was not able to optimize feature embedding to enforce higher similarity for intra-class and lower similarity for inter-class samples. To enforce their similarity, the additive angular margin loss (AAM-Softmax)\cite{deng2019arcface} concept has appeared by giving some margin $m$ for the different classes. Additive angular margin loss function normalize weight $||W_j|| = 1$ with $l_2$ normalization. AAM-Softmax is represented as follows: 

    \begin{equation}\label{eq2}
        L(x_i) =  - \log ( \frac{e^{s\:cos({\theta}_{y_i} + m)}}{e^{s \: cos({\theta}_{y_i} + m)} + \Sigma_{j=1, j \neq y_i}^{N} e^{s \: cos{\theta}_{j}} })
    \end{equation}

    $s$ is the scale parameter that we need to scale the whole embedding features space size, and $m$ is a hyperparameter for margin. High values for $m$ make initial training difficult and slow down the whole training process, however, it helps to increase embedding feature space diversity. 

    We additionally used subcenter technique to keep temporal center information while normalize weight $||W_j|| = 1$ with $l_2$ normalization. When we normalize AAM-Softamx with subclasses, the dimension of the weight parameter changed to $W \in {\rm I\!R} ^{512 \times N \times K}$, and we use a new cosine similarity that is processed with subclass-wise max pooling.

\subsection{Training protocols}
    
    Our training methods consist of two stages. At the initial stage, the margin of AAM-Softmax was set to 0.3 with two subcenters. We randomly cropped utterances into 6 seconds and selected 384 segments for a mini-batch. The SGD updated the parameters with a momentum of 0.9 and a weight decay of 0.001. For the learning rate scheduler, we utilized cosine annealing with restarts \cite{loshchilov2016sgdr}. The initial iteration of cosine was set to 1 epoch, and each iteration was doubled after every training cycle. We started training with a maximum learning rate of 0.02 and decayed with a factor of 0.8 at every restart. The minimum learning rate was fixed at 5e-6. The model checkpoints were created at the end of the cosine cycle and examined with VoxCeleb1-O to find the best one.

    The second stage was large margin fine-tuning \cite{thienpondt2020idlab} for the best checkpoints mentioned above. We increased the margin from 0.3 to 0.5 or 0.6 and decreased the maximum learning rate to 1e-4. The learning rate decayed with the cosine annealing scheduler and restarted every 11,000 training steps. Unlike the original implementation, we did not increase the length of input segments and applied no sampling strategy. For the fine-tuning stage, it was important to examine the validation metrics frequently and stop the training before the divergence. We found that most of the model diverges within a few thousand parameter updates before the learning rate restarts. Detailed training configurations are listed in Table \ref{tab:sys_arch}.
    
    \begin{table}[t]
      \caption{Training protocols and scoring methods for single and fusion systems (\textbf{TC}: number of training cycles, \textbf{LMF}: large margin fine-tuning)}
      \label{tab:sys_arch}
      \centering
      \begin{tabular}{clccc}
        \toprule
         & \textbf{Neural network} & \textbf{TC} & \textbf{LMF} & \textbf{Scoring} \\
        \midrule
        \textbf{S1} & ResNet34-st1112 & 4 & - & AS-Norm \\
        \textbf{S2} & ResNet34-st1121 & 5 & - & AS-Norm \\
        \textbf{S3} & ResNet34-st1121 & 6 & 0.6 & MSA \\
        \textbf{S4} & ResNetSE101 & 4 & - & AS-Norm \\
        \textbf{S5} & ResNetSE101 & 4 & 0.5 & AS-Norm \\
        \textbf{S6} & ResNetSE101 & 5 & 0.5 & AS-Norm \\
        \textbf{S7} & ResNetSE101 & 4 & 0.5 & MSA \\
        \bottomrule
        \textbf{Fusion} & Weighted sum \\
      \end{tabular}
    \end{table}

\begin{table*}[t]
  \caption{Evaluation results on development sets}
  \label{tab:sys_dev}
  \centering
  \begin{tabular}{ccccccccc}
    \toprule
     & \multicolumn{2}{c}{\textbf{Vox1-O}} & \multicolumn{2}{c}{\textbf{VoxSRC20-dev}} & \multicolumn{2}{c}{\textbf{VoxSRC21-val}} & \multicolumn{2}{c}{\textbf{VoxSRC22-dev}} \\
    \textbf{} & \textbf{EER(\%)} & \textbf{DCF$_{0.05}$} & \textbf{EER(\%)} & \textbf{DCF$_{0.05}$} & \textbf{EER(\%)} & \textbf{DCF$_{0.05}$} & \textbf{EER(\%)} & \textbf{DCF$_{0.05}$} \\
    \midrule
    \textbf{S1} & 0.568 & 0.0365 & 2.156 & 0.1016 & \textbf{2.075} & 0.1074 & 1.573 & 0.1082 \\
    \textbf{S2} & 0.521 & 0.0284 & 2.135 & 0.1001 & 2.174 & 0.1102 & 1.555 & 0.1011 \\
    \textbf{S3} & 0.627 & 0.0336 & 2.339 & 0.1187 & 2.399 & 0.1290 & 1.627 & 0.0988 \\
    \textbf{S4} & 0.484 & 0.0310 & 2.054 & 0.1005 & 2.137 & 0.1075 & 1.550 & 0.1047 \\
    \textbf{S5} & 0.505 & 0.0306 & \textbf{2.044} & 0.0996 & 2.138 & \textbf{0.1029} & 1.542 & 0.1036 \\
    \textbf{S6} & \textbf{0.430} & \textbf{0.0245} & 2.057 & \textbf{0.0966} & 2.201 & 0.1154 & 1.570 & 0.1050 \\
    \textbf{S7} & 0.632 & 0.0365 & 2.202 & 0.1120 & 2.257 & 0.1222 & 1.593 & 0.0970 \\
    \bottomrule
    \textbf{Fusion} & & & & & & & \textbf{1.370} & \textbf{0.0863} \\
  \end{tabular}
\end{table*}
    
\subsection{Score estimation}

    For the verification test, the speaker embeddings were extracted in 512-dimension with length normalization. We tested the adaptive symmetric normalization (AS-Norm) \cite{zhao2021speakin, xiang2020xx205, wang2020dku, matejka2017analysis} and the matrix score average (MSA) \cite{heo2020clova, torgashov2020id, zhang2021beijing} for the scoring method. For AS-Norm, we utilized the 5,994 speakers in the development set of VoxCeleb2 as imposter cohorts. Then the similarity score was calibrated by the top 100 imposters with symmetric normalization.
    
    For MSA, We split utterances into five segments with the allowance of overlaps to formulate a scoring matrix with a size of $\mathbb{R} ^{(5 \times 5)}$ for each trial. Padding was applied if the length of the utterance was shorter than 6 seconds. We averaged the 25 scores to represent the trial score.

    We used the VoxSRC22-dev to select the best scoring method for each system. The final composition of the single system is given in Table \ref{tab:sys_arch}.

\subsection{System fusion}

    We applied the late fusion technique that ensemble all individual systems with prediction scores. Our strategy was to utilize logistic regression to find the contribution of each system that optimizes the final metric of VoxSRC22-dev. This method slightly improved the result than the average sum of prediction scores.

\section{Results}
We evaluated the systems with the Equal Error Rate (EER) and the minimum Detection Cost Function (DCF) with $P_{traget}=0.05$. We analyzed the results with the DCF as the primary metric and the EER as the supplementary metric. Since the DCF has been used with different $P_{traget}$ values in previous studies, it is required to compare the results in the EER. The system evaluation results for the development data are shown in Table \ref{tab:sys_dev}, and the final submission results are shown in Table \ref{tab:sys_eval}.

\begin{table}[th]
  \caption{Submission results}
  \label{tab:sys_eval}
  \centering
  \begin{tabular}{ccc}
    \toprule
       & \textbf{EER(\%)} & \textbf{DCF$_{0.05}$} \\
    \midrule
    \textbf{S1} & - & - \\
    \textbf{S2} & 3.564 & 0.1938 \\
    \textbf{S3} & - & - \\
    \textbf{S4} & 3.424 & 0.1911 \\
    \textbf{S5} & - & - \\
    \textbf{S6} & - & - \\
    \textbf{S7} & - & - \\
    \bottomrule
    \textbf{Fusion} & 2.912 & 0.165 \\
  \end{tabular}
\end{table}

\bibliographystyle{IEEEtran}

\bibliography{mybib}

\end{document}